# Analytical Model for Light Modulating Impedance Spectroscopy (LIMIS) in All-Solid-State p-n Junction Solar Cells at Open-Circuit

Osbel Almora[1,2,3]*, Daniel Miravet[4], Gebhard J. Matt[1], Germà Garcia-Belmonte[3] and Christoph J. Brabec[1]

*[1]Institute of Materials for Electronics and Energy Technology (i-MEET), Friedrich-Alexander-Universität Erlangen-Nürnberg, 91058 Erlangen, Germany;*

*[2]Erlangen Graduate School in Advanced Optical Technologies (SAOT), Friedrich-Alexander-Universität Erlangen-Nürnberg, 91052 Erlangen, Germany*

*[3]Institute of Advanced Materials (INAM), Universitat Jaume I, 12006 Castelló, Spain*

*[4]Centro Atómico Bariloche, CNEA, CONICET, Río Negro, Argentina*

*osbel.almora@fau.de

## Abstract

Non-circuit theory drift-diffusion numerical simulation of standard potentiostatic impedance spectroscopy (IS) is a well-known strategy for characterization of materials and electronic devices. It implies the time-dependent solutions from the continuity and Poisson's equations under small perturbation of the bias boundary condition at the electrodes. But in the case of photo-sensitive devices a small light perturbation can be also taken modulating the generation rate along the absorber bulk. In that focus, this work approaches a set of analytical solutions for the signals of IS and intensity modulated photocurrent and photovoltage spectroscopies, IMPS and IMVS respectively, from one-sided p-n junction solar cells at open-circuit. Subsequently, a photo-impedance signal named "light intensity modulated impedance spectroscopy" (LIMIS=IMVS/IMPS) is analytically simulated and its difference with respect to IS suggests a correlation with the surface charge carrier recombination velocity. This is an illustrative result and starting point for future more realistic numerical simulations.

The concept of impedance as a transfer function in a form of a ratio between two complex magnitudes has been widely tackled, since its introduction by Heaviside.[1] The most typical application is for the study of electrical current response to small voltage perturbation, as in the standard potentiostatic impedance spectroscopic (IS) where the impedance itself $Z$ has units of Ohms.[2] The IS is a well-known and stablished characterization technique for the evaluation of the resistive, capacitive and inductive features of materials and electronic devices. Particularly, on photovoltaic solar cells IS typically informs on the recombination modes,[3] the doping densities,[4-5] deep defect levels[6] and the density of states.[7] One of the most common practices is to measure potentiostatic IS at open-circuit (OC) condition under a series of steady-state illumination intensities. In this way, from numerical circuit theory calculations the recombination resistance $R_{rec}$, chemical capacitance $C_{\mu}$ and characteristic lifetimes $\tau$ can be evaluated, among other parameters.[3, 8-9] Also drift-diffusion numerical calculations have been reported solving the time-dependent continuity and Poisson's equations under small perturbation of the bias boundary condition at the electrodes.[10-13]

Alternatively, in photo-sensitive samples the current or voltage responses to small light intensity perturbations can be also studied, which are the cases of the intensity modulated photocurrent spectroscopy (IMPS)[14-25] and the intensity modulated photovoltage spectroscopy (IMVS),[20-21, 26-28] respectively. IMVS and IMPS individually characterize the current and voltage responsivities $\Psi_J$ and $\Psi_V$, respectively. Hence, similarly to IS, we can take the ratio IMVS/IMPS to define a light intensity modulated impedance spectroscopy (LIMIS). This relation was first introduced by Song & Macdonald[29] who measured the spectra on n-Si in KOH solution, validating the transfer function by Kramers-Kronig transformation. Later Halme[30] applied the concept on dye sensitized solar cells,  concluding the approximate equivalence between IS and LIMIS. Furthermore, in our simultaneous and complementary study[8] we measured LIMIS in all-solid-state silicon, organic and perovskite solar cells, and numerically simulated LIMIS spectra with circuit-theory, showing qualitative and quantitative differences between IS and LIMIS. These differences were showed to imply corrections to the carrier lifetimes evaluations and the *dc* empirical Shockley equation.

Understanding the difference between the transfer functions from IS and LIMIS demands a model able to reproduce and explain the experimental patterns. Thus, the accurate solving of the drift-diffusion equations would possibly be the best resource, requiring the use of numerical methods to reproduce the frequency-dependent signals around the OC steady-state under *dc* illumination. In the case of LIMIS, or individually IMPS and IMVS, the complete development of the time dependent numeric solutions is still in early phases[10] and such task is beyond the scope of this paper. Instead, the focus is set here on the analytical solution for the particular case of one-sided abrupt p-n junction thin film solar cells.

In this article we further analyze the LIMIS concept at OC and solve the drift-diffusion equations in an analytical approximation for the one-sided p-n junction solar cells that suggest a correlation between the difference among IS and LIMIS and the surface recombination velocity. These analytical results complement the typical circuit theory studies and set a starting point for future drift-diffusion simulations.

Introducing the formalism, with the notation in table S1, let's first consider a sinusoidal $\widetilde{V}(t)$ small perturbation applied to a generic sample at steady-state voltage $\overline{V}$. Then the current response around the steady-state value $\bar{J}(\overline{V})$ may be $\phi$ phase shifted, as $\tilde{J} = |\tilde{J}|\exp[-i\phi]$ resulting the impedance as

$$Z(\omega) = \frac{\widetilde{V}(t)}{\tilde{J}(t)} = \frac{|\widetilde{V}|}{|\tilde{J}|}\exp[i\phi] \tag{1}$$

Now, instead, the perturbation can be done by a light source in photo-sensitive samples. Then, a small perturbation $\widetilde{P}_{in}(t) = |\widetilde{P}_{in}|\exp[i\ \omega t]$ can be added to the given *dc* incident light power density $\overline{P}_{in}$. Upon this perturbation, both current and voltage signals can be recorded, so the current responsivity transfer function is

$$\Psi_J(\omega) = \frac{\tilde{J}}{\widetilde{P}_{in}} = \frac{|\tilde{J}|}{|\widetilde{P}_{in}|}\exp[i\phi_J] \tag{2}$$

and, at OC ($J = 0$) the modulated photovoltage signal $\widetilde{V}_{oc} = |\widetilde{V}_{oc}|\exp[i\phi_V]$ give the voltage responsivity transfer function as

$$\Psi_V(\omega) = \frac{\widetilde{V}_{oc}}{\widetilde{P}_{in}} = \frac{|\widetilde{V}_{oc}|}{|\widetilde{P}_{in}|}\exp[i\phi_V] \tag{3}$$

Equations (2) and (3) define by themselves IMPS and IMVS, respectively. These techniques have been earlier introduced[14-17, 23, 27, 31] and there have been recent studies on photovoltaic solar cells.[21-22, 24, 28, 32] Now, as discussed before,[8] it can be advantageous to combine IMPS and IMVS instead or complementary to their individual analyses. Therefore we obtain the "light intensity modulated impedance spectroscopy" (LIMIS) as

$$Z_\Psi(\omega) = \frac{\Psi_V}{\Psi_J} = \frac{|\tilde{V}_{oc}|}{|\tilde{J}|}\exp[i(\phi_V - \phi_J)] = |Z_\Psi|\exp[i\phi_\Psi] \quad (4)$$

The individual expressions for the modulated currents and voltages are further introduced in equations (S1)-(S8) in Section S1.1. In Section S1.2. we present the continuity equation (S9) including the drift and diffusion terms, the Poisson's equation, the drift-diffusion currents and the boundary conditions (S10) for the electrostatic potential $\varphi$ and the current in the assumption of ohmic contacts. In this formalism we highlight that IS and LIMIS are different regarding "where" the perturbation is included. In IS $\tilde{V}$ directly affects the $\varphi$ boundary condition, which defines the electric field $\xi$ after the Poisson's equation. Later, its effect will be particularly related with recombination and its influence in the space charge region in the continuity equation.

On the other hand, the perturbed term in LIMIS is directly affecting the continuity equation via the generation rate $G$ along the effective absorber layer bulk section. Assuming a light intensity independent incident light spectrum, $G$ can be expressed in *dc* and *ac* real terms as

$$G(t) = \bar{G} + \tilde{G}\exp[i\omega t] = \frac{\Psi_{sc}}{q\,L}\left(\bar{P}_{in} + \tilde{P}_{in}\exp[i\omega t]\right) \quad (5)$$

where $L$ here is the effective absorber layer thickness where the current is integrated and $\Psi_{sc}$ is the photo-current responsivity at short-circuit that depends on the incident light spectrum, the absorption coefficient and geometry of the absorbing materials. Note that $\tilde{G} = |\tilde{G}|$ and $\tilde{P}_{in} = |\tilde{P}_{in}|$ are the perturbation, and similarly to $\tilde{V}$, we will omit the modulus notation in the next. Also note that only in thin film devices (5) can

be approached to an space independent constant $G$, otherwise the Beer–Lambert law should be considered. Subsequently, the inclusion of $\tilde{G}$ in the continuity equation defines the diffusion currents out of the space charge region, or in situations for low field effects. This can be particularly significant for the current boundary condition.

Keeping this in mind, in Sections S1.1-8 the analytical solution of the charge carrier concentrations around OC under $\tilde{V}$ and $\tilde{G}$ perturbations, for IS and LIMIS respectively, are presented. The main idea is to structure the solutions in the form

$$n(t) = \bar{n}_0 + \tilde{n} \exp[i\omega t] \tag{6}$$

where $\bar{n}_0 = \bar{n} + n_0$, $n_0$ is the dark equilibrium concentration, $\bar{n}$ the steady-state over-equilibrium-concentration (under *dc* bias and/or illumination) and $\tilde{n}$ is the complex amplitude response to $\tilde{V}$ or $\tilde{G}$ which includes the phase shift $\phi_n$ , i.e., $\tilde{n} = |\tilde{n}|\exp[i\phi_n]$. The current boundary condition was taken as (S11) in the form of ohmic contact selectivity with negligible drift current, where minority carriers recombine with surface recombination velocity $S_r$. No significant difference was considered between IS and LIMIS regarding the $S_r$ constrain. On the other hand, the potential boundary condition was chosen as the depletion approximation (S12) expressing how the different measurement ways ideally affect the charge carrier distributions and hence the energy diagram. This is summarized in **Figure 1**. For IS the $\tilde{V}$ small perturbation changes the depletion region width $w$ and creates small charge carrier gradients around the steady-state OC distribution. For the IMVS the $\tilde{G}$ small perturbation also changes $w$, but without gradients, so the charge profile is flat all the time. Finally, for IMPS no change of $w$ is assumed and the opposite gradient direction takes place.

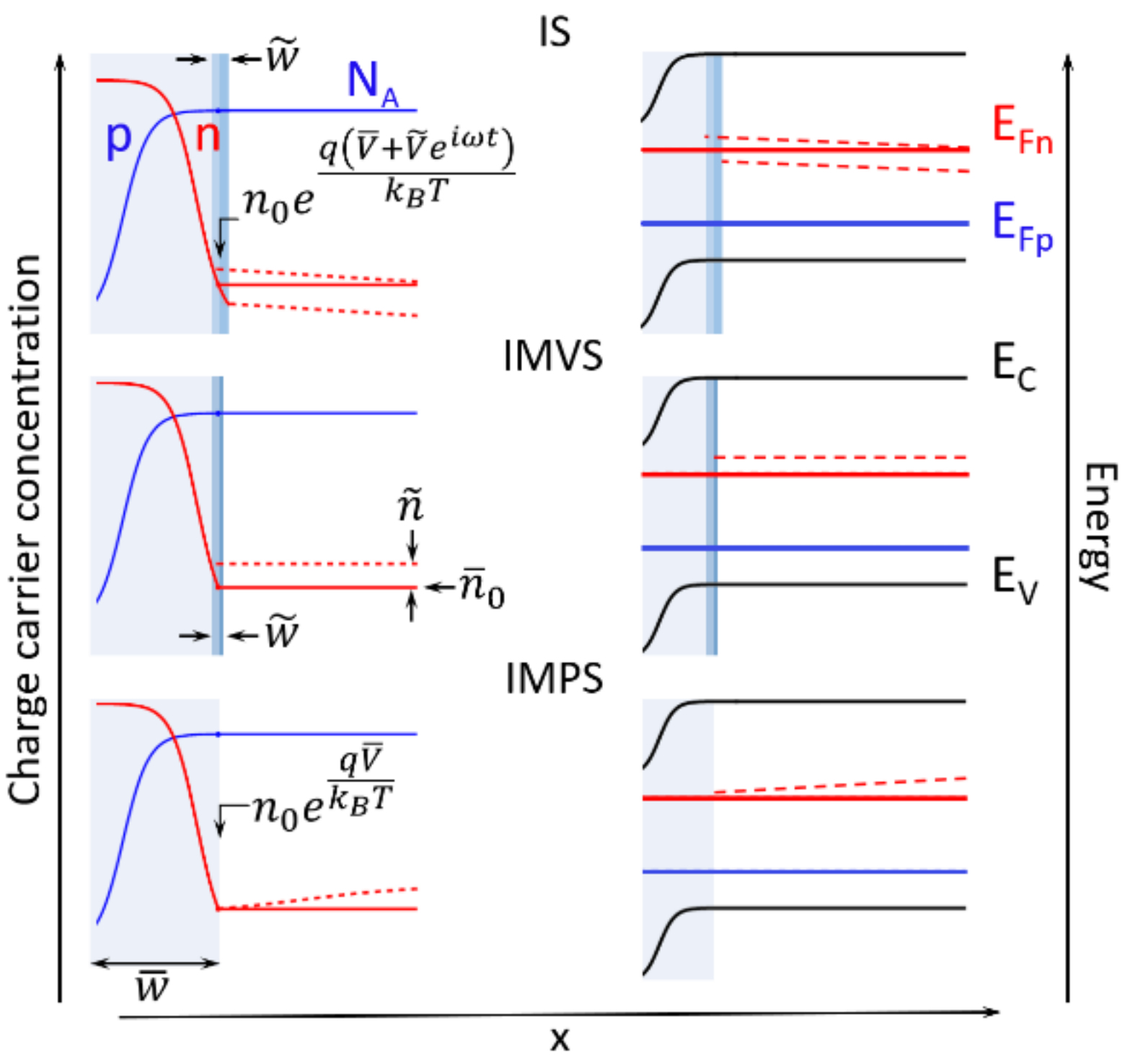

**Figure 1.** Charge carrier (left) and Energy diagram (right) for the perturbations around open circuit for IS, IMPV and IMPS, as indicated. The dashed lines represent the perturbed quantities. $E_C$ and $E_V$ are de conduction and valence bands minima and maxima, respectively, $\bar{w}$ and $\tilde{w}$ are likewise the *dc* and *ac* parts of the depletion region width.

Accordingly, as deducted in section S1.5. the impedance from IS at OC can be written as

$$Z(\omega) \cong \frac{2k_B T}{q^2 \bar{L}_d \bar{G}} \frac{1}{\tilde{\gamma}\sqrt{1 + i\dfrac{\omega}{\omega_0}}} \tag{7}$$

Where $\bar{L}_d = \sqrt{D/\omega_0}$ is the diffusion length, $D$ the diffusion coefficient, $\omega_0$ a characteristic recombination frequency as (S19.b) and $\tilde{\gamma}$ is a surface recombination factor as (S24.c).

Furthermore, in sections S1.6,7 the *ac* voltage and current responsivities were deducted as

$$\Psi_V(\omega) = 2\frac{k_B T}{q\bar{G}}\frac{1}{\left(1 + i\frac{\omega}{\omega_0}\right)} \tag{8.a}$$

$$\Psi_J(\omega) = \frac{q\bar{L}_d}{\sqrt{1 + i\frac{\omega}{\omega_0}}}\tilde{\gamma}\left(1 - \tilde{\delta}\right) \tag{8.b}$$

with the complex diffusion length $\tilde{L}_d$ as (S21.b), the factor $\tilde{\gamma}$ as (S24.c) and

$$\tilde{\delta} = S_r \frac{\left(e^{\frac{\bar{w}}{\tilde{L}_d}} - e^{\frac{L-\bar{w}}{\tilde{L}_d}}\right)}{\left(\frac{D}{\tilde{L}_d} + S_r + \left(\frac{D}{\tilde{L}_d} - S_r\right)e^{\frac{L}{\tilde{L}_d}}\right)} \tag{8.c}$$

Note that (8) shows how the $\omega$-dependency towards higher frequencies both are expected to behave with arc-shape-type spectra since $\Psi_V \propto \left(1 + i\frac{\omega}{\omega_0}\right)^{-1}$ and $\Psi_J \propto \left(1 + i\frac{\omega}{\omega_0}\right)^{-1/2}$.Interestingly, (8) also suggest that at the low frequency limit $\Psi_V \propto \bar{G}^{-1}$ but $\Psi_J$ should be nearly $\bar{G}$-independent. Particularly, the low frequency limit from IMVS spectra can be used to straightforwardly evaluate the ideality factor $m$ as demonstrated in (S28) resulting

$$\Psi_V \cong m\frac{k_B T}{q\bar{G}} \tag{8.d}$$

In our simultaneous work[8] we showed the agreement of (8.d) with $m$ obtained by IS and photocurrent-photovoltage curves of silicon, organic and perovskite solar cells.

Subsequently, from (8.a,b) it is easy to obtain our LIMIS transfer function as

$$Z_\Psi(\omega) \cong \frac{2k_B T}{q^2\bar{L}_d\bar{G}}\frac{1}{\tilde{\gamma}\sqrt{1 + i\frac{\omega}{\omega_0}}\left(1 - \tilde{\delta}\right)} \tag{9}$$

Expression (9) for LIMIS is different to (7) for IS in only a factor, so we can obtain the IS-LIMIS normalized impedance as

$$\Delta Z_{\Psi} = \frac{Z_{\Psi} - Z}{Z} = \frac{\tilde{\delta}}{1 - \tilde{\delta}} \tag{10}$$

which is directly proportional to $S_r$, as evident in (8.c) and more explicitly in (S35), meaning that $\Delta Z_{\Psi}$ as (10) may inform on the surface recombination at the electrodes.

The model was used to simulate IS and LIMIS spectra from a p-n junction solar cell, whose full characterization can be found in our simultaneous work.[8] **Figure 1** shows the experimental data and simulations in Nyquist plot representation for the simulation parameters in Table S2.

In a first example **Figure 1**a shows two spectra where IS and LIMIS are very similar, only the latter shows a slightly higher resistance. Here a same extra series resistance $R_s$ was added to the real parts of both impedances. This pattern can be well simulated towards low frequencies by our model. Also in that figure, with dashed lines we illustrate the role of increasing recombination via reducing the Shockley-Read-Hall lifetime $\tau$ of increasing the band-to-band recombination coefficient $\beta$. This result in the decrease of the impedances and the sign inversion of $\Delta Z_{\Psi}$, meaning that IS may deliver more impedance than LIMIS, an already observed feature.[8]

In **Figure 1**b we highlight the resistances $R_{IS}$ and $R_{LIMIS}$ corresponding to the IS and LIMIS spectra, respectively. In this regard the situation is not so different to **Figure 1**a, being $R_{LIMIS}$ not too higher than $R_{IS}$. But differently, the LIMIS spectrum is significantly right shifted in what looks like a series resistance effect at higher frequencies. The understanding of this feature is of particular difficulty due to the experimental limitations for measuring IMPS and IMVS and the reported more complex spectra shapes.[8] Also it cannot be explained by our model so an extra term $Z_s{}'$ was added to the $\mathrm{Re}[Z_{\Psi}(\omega)]$ in addition to the $R_s$, in common with the IS spectra. Then **Figure 1**c displays the situation where $R_{LIMIS}$ is no longer higher than $R_{IS}$, but still the $Z_s{}'$ effect delivers a larger low frequency impedance from LIMIS with respect to IS.

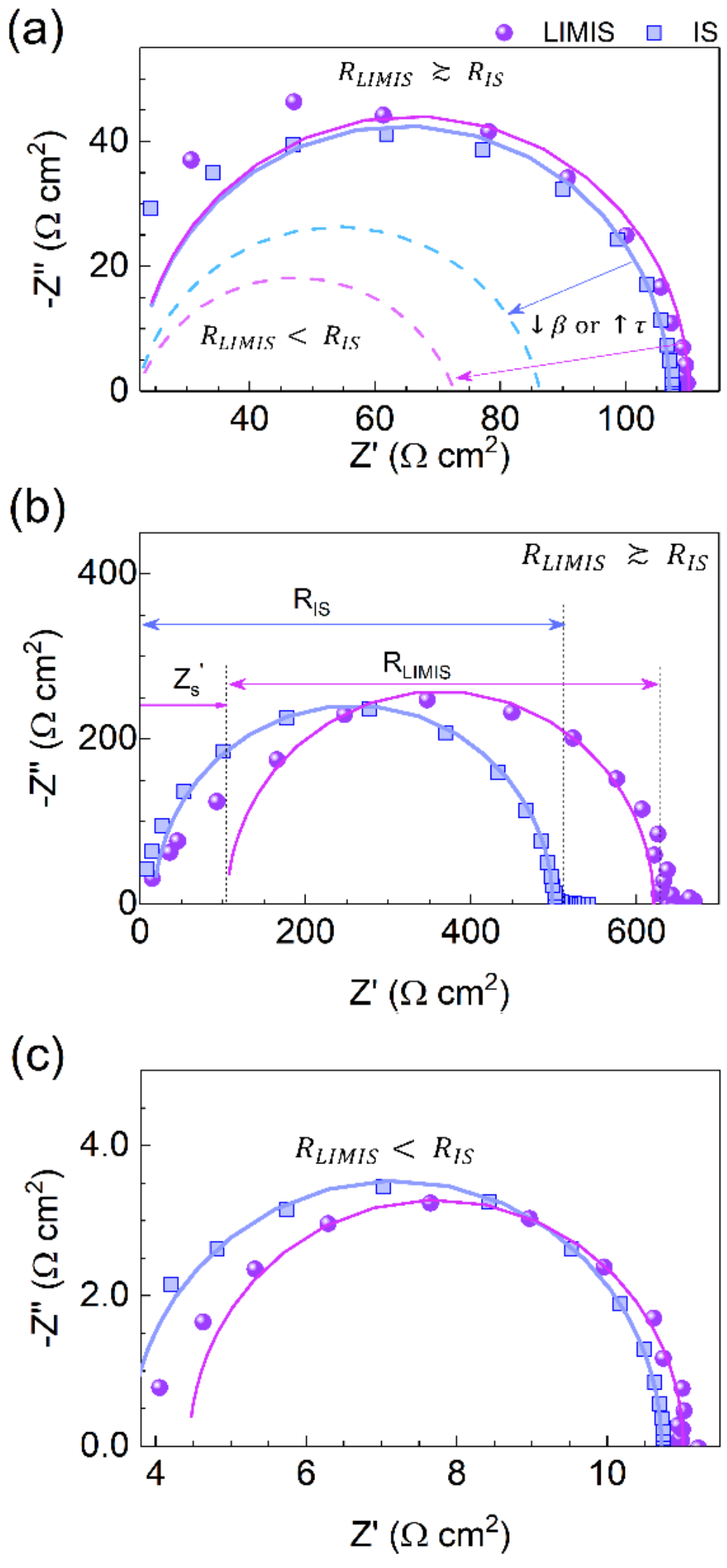

**Figure 2.** Experimental spectra (dots) and our model simulations (lines) for IS, and LIMIS characterization of a standard p-n junction silicon solar cell. The simulation parameters are summarized in Table S2.

In summary, the time-dependent analytical solutions for the continuity equation considering diffusion currents in a one-sided p-n junction solar cell were found around the open circuit steady-state upon small bias and light perturbations. This allowed to show analytical expressions for the transfer functions of the potentiostatic impedance IS and the photo-impedance LIMIS. The difference between LIMIS and IS respective impedances, resulted directly proportional to the interface recombination velocity. The model was used to simulate experimental spectra from a silicon solar cell showing a good agreement, mainly at lower frequencies. Extra experimental features like a series resistances-like right-shifting in the LIMIS Nyquist plots point out the limitations of the model towards high frequencies, suggesting the need for more realistic and numerical approaches.

## Conflicts of interest

There are no conflicts to declare.

## Acknowledgements

We thank Ministerio de Ciencia, Innovación y Universidades of Spain under project (MAT2016-76892-C3-1-R). O.A. acknowledges the financial support from the VDI/VD Innovation + Technik GmbH (Project-title: PV-ZUM) and the SAOT funded by the German Research Foundation (DFG) in the framework of the German excellence initiative.

## AUTHOR INFORMATION

### Corresponding Author

*E-mail: osbel.almora@fau.de

### ORCID

Osbel Almora: 0000-0002-2523-0203

Daniel Miravet: 0000-0002-2908-4645

# *Supporting Information:*

# Analytical Model for Light Modulating Impedance Spectroscopy (LIMIS) in All-Solid-State p-n Junction Solar Cells at Open Circuit

Osbel Almora[1,2,3]*, Daniel Miravet[4], Gebhard J. Matt[1], Germà Garcia-Belmonte[3] and Christoph J. Brabec[1]

[1]*Institute of Materials for Electronics and Energy Technology (i-MEET), Friedrich-Alexander-Universität Erlangen-Nürnberg, 91058 Erlangen, Germany;*

[2]*Erlangen Graduate School in Advanced Optical Technologies (SAOT), Friedrich-Alexander-Universität Erlangen-Nürnberg, 91052 Erlangen, Germany*

[3]*Institute of Advanced Materials (INAM), Universitat Jaume I, 12006 Castelló, Spain*

[4]*Centro Atómico Bariloche, CNEA, CONICET, Río Negro, Argentina*

*osbel.almora@fau.de

**Table S1:** List of acronyms, symbols and abbreviations

| | |
|---|---|
| 2D | Two-dimensional |
| 3D | Three-dimensional |
| *ac* | Alternating current (mode) |
| $\beta$ | Radiative recombination coefficient ($cm^3 \cdot s^{-1}$) |
| *b* | Power law for the relation *dc* resistance vs. incident light intensity |
| *c* | Speed of light in vacuum (299 792 458 $m \cdot s^{-1}$) |
| $C$ | Capacitance ($F \cdot cm^{-2}$) |
| $C^*$ | Complex capacitance ($F \cdot cm^{-2}$) |
| $C_{bulk}$ | Bulk capacitance ($F \cdot cm^{-2}$) |
| CB | Conduction band |
| $C_{diff}$ | Diffusion layer cap. ($F \cdot cm^{-2}$) |
| $C_{dl}$ | Depletion layer cap. ($F \cdot cm^{-2}$) |
| $C_g$ | Geometric capacitance ($F \cdot cm^{-2}$) |
| $C_{Hf}$, $C_{Lf}$ | High and low frequencies capacitances, respectively, from IS and LIMIS spectra ($F \cdot cm^{-2}$) |
| $C_{Hl}$ | Helmholtz layer capacitance ($F \cdot cm^{-2}$) |
| $C_{\mathrm{IS}}$ | Capacitance from IS ($F \cdot cm^{-2}$) |
| $C_{\Psi}$ | Capacitance from LIMIS ($F \cdot cm^{-2}$) |
| $C_{\mu}$ | Chemical capacitance ($F \cdot cm^{-2}$) |
| *dc* | Direct current (mode) |
| DD | Drift-diffusion |
| DFT | Density function theory |
| $\mathfrak{D}$ | Electric displacement ($C \cdot cm^{-2}$) |
| $\tilde{\delta}$ | Complex *ac* difference factor between IS and LIMIS |
| $D$ | Diffusion coefficient ($cm^2 \cdot s^{-1}$) |
| $D_n$, $D_p$ | Diffusion coefficient for electrons and holes, respectively ($cm^2 \cdot s^{-1}$) |
| $\Delta Z_{\Psi}$ | Complex normalized photo-impedance difference LIMIS-IS |
| $\Delta Z_{\Psi}'$ | normalized difference of real parts of photo-impedance LIMIS-IS |
| $\varepsilon$ | Dielectric constant |
| $\varepsilon_0$ | Vacuum permittivity ($8.85 \times 10^{14}$ $F \cdot cm^{-1}$) |
| $E$ | Energy (eV or J) |
| EC | Equivalent circuit |
| $E_C$ | Conduction band minimum energy level (eV) |
| $E_{Fn}$, $E_{Fp}$ | Quasi-Fermi level of electrons and holes, respectively (eV) |
| $E_g$ | Band-gap energy (eV) |
| $E_i$ | Intrinsic energy level (eV) |
| $E_V$ | Valence band maximum level (eV) |
| ESL | Electron selective layer |
| ETL | Electron transport layer |
| ETM | Electron transport material |
| EQE | External quantum efficiency |
| $f$ | Frequency (Hz) |
| $f_{\tau}$ | Characteristic frequency (Hz) |
| FA | Formamidinium |
| $FF$ | Fill factor |
| $\tilde{\gamma}$, $\tilde{\gamma}_1$ | Complex *ac* surface recombination factors |
| $G$ | Generation rate ($cm^{-3} \cdot s^{-1}$) |
| $G_0$ | Generation rate at $x = 0$ ($cm^{-3} \cdot s^{-1}$) |
| $\tilde{G}$ | Real *ac* perturbation generation rate amplitude ($cm^{-3} \cdot s^{-1}$) |
| $\bar{G}$ | Real *dc* generation rate ($cm^{-3} \cdot s^{-1}$) |
| $h$ | Planck's constant ($6.626 \times 10^{-34}$ $J \cdot s$) |
| HSL | Hole selective layer |
| HTL | Hole transport layer |
| HTM | Hole transport material |
| $i$ | Imaginary number ($\sqrt{-1}$) |
| IMPS | Intensity modulated photocurrent spectroscopy |
| IMVS | Intensity modulated photovoltage spectroscopy |
| IHys | Inverted hysteresis |
| IS | Impedance spectroscopy |
| $J$ | Current density ($A \cdot cm^{-2}$) |
| $\tilde{J}$ | Complex *ac* current density signal amplitude ($A \cdot cm^{-2}$) |
| $\bar{J}$ | Real *dc* current density ($A \cdot cm^{-2}$) |
| $J_n$ | Electron current density ($A \cdot cm^{-2}$) |
| $J_p$ | Holes current density ($A \cdot cm^{-2}$) |
| $J_{ph}$ | Photocurrent density ($A \cdot cm^{-2}$) |
| $J_s$ | Reverse bias diode dark saturation current density ($A \cdot cm^{-2}$) |
| $J_{sc}$ | Short-circuit current density ($A \cdot cm^{-2}$) |
| $J - V$ | Current density-voltage characteristic (plane) |
| $k_B$ | Boltzmann constant ($1.38 \times 10^{-23}$ $J \cdot K^{-1}$) |
| $\lambda$ | Photon wavelength (nm) |
| $L$ | Distance between selective contacts/ Effective distance for current integration (cm) |
| $L_{bulk}$ | Thickness of the absorber bulk layer (cm) |
| $L_d$ | Diffusion length (cm) |
| $\tilde{L}_d$ | Complex *ac* diffusion length signal amplitude (cm) |
| $\bar{L}_d$ | Real *dc* diffusion length (cm) |
| $L_D$ | Debye length (cm) |
| LIMIS | Light intensity modulated impedance spectroscopy |
| LIMTAS | Light intensity modulated thermal admittance spectroscopy |
| $\mu$ | Electronic mobility ($cm^2 \cdot V^{-1} \cdot s^{-1}$) |
| $\mu_n$, $\mu_p$ | Electrons and holes mobilities, respectively ($cm^2 \cdot V^{-1} \cdot s^{-1}$) |
| $m$ | Diode ideality factor |
| $m_C$ | Capacitance ideality factor |
| $m_\Psi$ | Photocurrent idelity factor |
| MA | Methylammonium |
| MAPI | $CH_3NH_3PbI_3$ |
| $n$ | Electron charge density/ Average minority carriers charge density ($cm^{-3}$) |
| $\tilde{n}$ | Complex *ac* average minority carriers charge density signal amplitude ($cm^{-3}$) |
| $\bar{n}$ | Real *dc* steady-state over-equilibrium minority carriers charge density ($cm^{-3}$) |
| $n_0$ | Real *dc* dark equilibrium minority carriers charge density ($cm^{-3}$) |
| $\bar{n}_0$ | Total real *dc* average minority carriers charge density ($cm^{-3}$) |
| $N_\mu$ | Effective total equilibrium charge density that contributes to chemical capacitance ($cm^{-3}$) |
| $N_A$ | Ionized fixed acceptor doping concentration ($cm^{-3}$) |
| $N_C$ | Effective density of states at the conduction band ($cm^{-3}$) |
| $N_{CV}$ | Average effective density of states at CB and VB ($cm^{-3}$) |
| $N_D$ | Ionized fixed donor doping concentration ($cm^{-3}$) |
| $N_{eff}$ | Effective concentration of fixed ionized species in the depletion zone: $N_D$ or $N_A$ ($cm^{-3}$) |
| $N_{ion}$ | Average concentration of ionized charge ($cm^{-3}$) |
| $N_\mu$ | Effective total equilibrium charge density that contributes to chemical capacitance ($cm^{-3}$) |
| $N_V$ | Effective density of states at the valence band ($cm^{-3}$) |
| $\omega$ | Angular frequency ($rad \cdot s^{-1}$) |
| $\omega_0$ | Characteristic recombination frequency ($rad \cdot s^{-1}$) |
| $\omega_\beta$ | Characteristic radiative recombination frequency ($rad \cdot s^{-1}$) |
| OC | Open-circuit |
| OrgSCs | Organic solar cells |
| $\varphi$ | Electrostatic Potential (V) |
| $\phi$ | Phase shift from IS (rad) |
| $\phi_J$ | Phase shift from IMPS (rad) |
| $\phi_n$ | Phase shift of the *ac* minority carriers signal amplitude (rad) |
| $\phi_V$ | Phase shift from IMVS (rad) |
| $\phi_\Psi$ | Phase shift from LIMIS (rad) |
| $p$ | Holes charge density ($cm^{-3}$) |
| $p_0$ | Real *dc* dark equilibrium minority carrier holes charge density ($cm^{-3}$) |
| PCE | Power conversion efficiency |
| PC | Photocurrent ($A \cdot cm^{-2}$) |
| $P_{in}$ | Light incident power ($W \cdot cm^{-2}$) |
| $\tilde{P}_{in}$ | Real *ac* light incident power perturbation amplitude ($W \cdot cm^{-2}$) |
| $\bar{P}_{in}$ | Real *dc* light incident power density ($W \cdot cm^{-2}$) |
| $\Psi_J$ | Current responsivity/ Complex current responsivity transfer function ($A \cdot W^{-1}$) |
| $\Psi_{J,dc}$ | Photo-current responsivity from the $dc\ J-V$ curve ($A \cdot W^{-1}$) |
| $\Psi_J'$, $\Psi_J''$ | Real and imaginary parts of $\Psi_J$, respectively ($A \cdot W^{-1}$) |
| $\Psi_{SC}$ | Real bias-independent current responsivity at SC ($A \cdot W^{-1}$) |
| $\Psi_V$ | Voltage responsivity/ Complex voltage responsivity transfer function ($V \cdot W^{-1} \cdot cm^2$) |
| $\Psi_{V,dc}$ | Photo-voltage responsivity from the $dc\ J-V$ curve ($V \cdot W^{-1} \cdot cm^2$) |
| $\Psi_V'$, $\Psi_V''$ | Real and imaginary parts of $\Psi_V$, respectively ($V \cdot W^{-1} \cdot cm^2$) |
| PSCs | Perovskite solar cells |
| PV | Photovoltaic |
| $q$ | Elementary charge ($1.6 \times 10^{-19}$ C) |
| $Q$ | Charge density ($C \cdot cm^{-2}$) |
| $\rho$ | Charge density ($C \cdot cm^{-3}$) |
| $R$ | Resistance ($\Omega \cdot cm^2$) |
| $R_{bulk}$ | Bulk resistance ($\Omega \cdot cm^2$) |
| $R_{dc}$ | $dc$ resistance from $J-V$ curve partial derivative ($\Omega\cdot cm^2$) |
| $R_{IS}$ | Resistance from IS ($\Omega\cdot cm^2$) |
| $r_{\Psi}$ | Photocurrent resistance factor |
| $R_{\Psi}$ | Resistance from LIMIS ($\Omega\cdot cm^2$) |
| $R_{\Psi,\mathrm{dc}}$ | LIMIS resistance from $dc$ $J-V$ curves ($\Omega\cdot cm^2$) |
| $R_T$ | Total $C$-coupled resistance ($\Omega\cdot cm^2$) |
| $R_{th}$ | Thermal recombination resistance ($\Omega\cdot cm^2$) |
| $R_s$ | Series resistance ($\Omega\cdot cm^2$) |
| $R_{sh}$ | Shunt resistance ($\Omega\cdot cm^2$) |
| SC | Short-circuit |
| SiSCs | Silicon solar cells |
| $S_r$ | Surface recombination velocity ($cm\cdot s^{-1}$) |
| $S_{rn}, S_{rp}$ | Surface recombination velocity of electrons and holes, respectively ($cm\cdot s^{-1}$) |
| $\tau$ | Lifetime/ Lifetime from TPV/ Non-radiative recombination lifetime/ Characteristic RC time constant from IS and LIMIS (s) |
| $\tau_{Hf}$, $\tau_{Lf}$ | High and low frequencies characteristic RC time constants from IS and LIMIS (s) |
| $\mathfrak{I}_{-1/2}$ | Fermi-Dirac 1/2 integral |
| $t$ | Time (s) |
| $T$ | Temperature (K) |
| $TAS$ | Thermal admittance spectroscopy |
| TPV | Transient photovoltage |
| $U$ | Recombination rate ($cm^{-3}\cdot s^{-1}$) |
| $\hat{v}_{1,2,3}$ | Unitary direction vectors |
| $V$ | Voltage (V) |
| $\tilde{V}$ | Real $ac$ voltage perturbation amplitude (V) |
| $\bar{V}$ | Real $dc$ voltage (V) |
| $V_{bi}$ | Built-in voltage (V) |
| VB | Valence band |
| $V_{\Psi}$ | Photocurrent resistance voltage |
| $V_{oc}$ | Open circuit voltage (V) |
| $\tilde{V}_{oc}$ | Complex $ac$ open circuit voltage signal amplitude (V) |
| $\bar{V}_{oc}$ | Real $dc$ open circuit voltage (V) |
| $w$ | Depletion layer width (cm) |
| $\tilde{w}$ | Complex $ac$ depletion layer width modulated amplitude (cm) |
| $\bar{w}$ | Real $dc$ depletion layer width (cm) |
| $\xi$ | Electric field ($V\cdot cm^{-1}$) |
| $x$ | Distance from the interface (cm) |
| $Z$ | Impedance/ Complex impedance transfer function from IS ($\Omega\cdot cm^2$) |
| $Z', Z''$ | Real and imaginary parts of $Z$, respectively ($\Omega\cdot cm^2$) |
| $Z_{\mathrm{T}}'$ | Total or low frequency limit of real part of impedance ($\Omega\cdot cm^2$) |
| $Z_{\Psi}$ | Photo-impedance from LIMIS/ Complex photo-impedance transfer function from LIMIS ($\Omega\cdot cm^2$) |

# S1. Theoretical deductions

### S1.1. Modulated magnitudes

The impedance is defined as the ratio between the oscillating voltage $\widetilde{V}$ and current $\tilde{J}$, being one a small perturbation and the other the response:

$$Z = \frac{\widetilde{V}}{\tilde{J}} \tag{S1}$$

In the potentiostatic variant of the impedance spectroscopy (IS) measurement, $\widetilde{V}$ is applied and the phase shifted $\tilde{J}$ is measured, for instance around open-circuit (OC) condition. Differently, in the light intensity modulated impedance spectroscopy (LIMIS) the light perturbation creates a phase shifted $\tilde{J}$ in the intensity modulated photo-current spectroscopy (IMPS) and another phase shifted open-circuit voltage response $\widetilde{V}_{oc}$ can be measured with the intensity modulated photo-voltage spectroscopy (IMVS).

In order to obtain theoretical expressions for $Z$, we approximate solutions of the charge density at open-circuit (OC). Such solutions, under a small sinus perturbation at angular frequency $\omega$, can be expressed as an average carrier concentration in the form

$$n(t) = \bar{n} + n_0 + \tilde{n}\exp[i\omega t] = \bar{n}_0 + \tilde{n}\exp[i\omega t] \tag{S2}$$

where $n_0$ is the dark equilibrium concentration, $\bar{n}$ is the steady-state over-equilibrium-concentration (under $dc$ bias and/or illumination) and $\tilde{n}$ is the complex amplitude response to the time $t$ varying perturbation which includes the phase shift corresponding to the response ($\tilde{n} = |\tilde{n}|\exp[i\phi_n]$). Note that, for simplicity, we analyze $n$ as a generic concentration of minority carriers at open circuit, or one of the

two in the one-side abrupt p-n junction approximation, being of particular focus the region within the diffusion length next to the depletion layer border.

In the potentiostatic IS case, a voltage small alternating current (*ac*) perturbation $\widetilde{V}$ is applied in addition to the direct current (*dc*) bias $\overline{V}$, so the applied voltage has the form

$$V(t) = \overline{V} + |\widetilde{V}| \exp[i\omega t] \tag{S3}$$

Note that here $\widetilde{V} = |\widetilde{V}|$ is a real amplitude (the one is set by the instrument), differently to $\widetilde{n}$ and the rest of the perturbed magnitudes. For simplicity we will use $\widetilde{V}$ in the following. Under the bias perturbation the current may similarly evolve as a modulated *ac* current $\tilde{J}$ added to the dc current $\bar{J}$ as

$$J(t) = \bar{J} + \tilde{J} \exp[i\omega t] \tag{S4}$$

In this case, similarly to $\widetilde{n}$, $\tilde{J}$ is a complex amplitude which carriers the phase shift information introduced by $\widetilde{n}$. Its relations can be approached by drift diffusion equations in the infinite mobility approximation and discarding second order small terms, so it results as[1]

$$\bar{J} \cong q \int_0^L \left[\bar{G} - \frac{\overline{n}_0}{\tau} - \beta \overline{n}_0{}^2\right] dx \tag{S5.a}$$

$$\tilde{J} \cong q \int_0^L \left[\tilde{G} - \tilde{n}\left(\frac{1}{\tau} + 2\beta\overline{n}_0 + i\omega\right)\right] dx \tag{S5.b}$$

Here $q$ is the elementary charge, $L$ is approximately the distance between electrodes but in practice the integral should be within the space charge region and diffusion lenghts, $\tau$ is the non-radiative surface/Shockley-Read-Hall (SRH) recombination lifetime, $\beta$ is the radiative recombination coefficient and $G$ is the generation rate, which is the perturbation for the LIMIS with $\bar{G}$ and $\tilde{G}$ as *dc* and *ac* real amplitudes, respectively, in the form

$$G(t) = \bar{G} + \tilde{G}\ \exp[i\omega t] = \frac{\Psi_{sc}}{q\ L}\left(\bar{P}_{in} + \tilde{P}_{in}\exp[i\omega t]\right) \qquad \text{(S6)}$$

Where $\Psi_{sc}$ is the photo-current responsivity at short-circuit that depends on the incident light spectrum, the absorption coefficient and geometry of the absorbing materials, and $\bar{P}_{in}$ and $\tilde{P}_{in}$ are the *dc* and *ac* real amplitudes of the incident light power, respectively. $L$ here is nearly the absorber layer thickness. Note that $\tilde{G} = |\tilde{G}|$ and $\tilde{P}_{in} = |\tilde{P}_{in}|$ are the perturbation, and similarly to $\tilde{V}$, we will omit the modulus notation in the next.

Under a perturbation like Equation (S6) a photo-voltage can be measured with $\bar{V}_{oc}$ and $\tilde{V}_{oc}$ as the *dc* and *ac* amplitudes, respectively, in the form

$$V_{oc}(t) = \bar{V}_{oc} + \tilde{V}_{oc}\exp[i\omega t] \qquad \text{(S7)}$$

Here $\tilde{V}_{oc}$ is a complex amplitude carrying the information of the phase shift, which can be deducted from standard equation and with the use of McLaurin series as[2]

$$\bar{V}_{oc} \cong \frac{E_g}{q} + 2\frac{k_B T}{q}\ln\left[\frac{\bar{n}_0}{N_{CV}}\right] \qquad \text{(S8.a)}$$

$$\tilde{V}_{oc} \cong 2\frac{k_B T}{q}\frac{\tilde{n}}{\bar{n}_0} \qquad \text{(S8.b)}$$

where $E_g$ is the absorber energy band-gap for the device, $k_B$ is the Boltzmann constant, $T$ is absolute temperature, and $N_{CV} = \sqrt{N_C N_V}$ is the square root of the average density of states at the conduction and valence bands, respectively, or the corresponding one in case of a one-side abrupt junction.

## S1.2. General equations for the numeric approach

The continuity equations for charge carrier concentrations of electrons ($n$) and holes ($p$) within the space between the electrodes[3]

$$\frac{\partial n}{\partial t} = G - U + \mu_n \, n \, \nabla\xi + \mu_n \, \xi \, \nabla n + D_n \, \nabla^2 n$$
$$\frac{\partial p}{\partial t} = G - U - \mu_p \, p \, \nabla\xi - \mu_p \, \xi \, \nabla p + D_p \, \nabla^2 p \qquad \text{(S9.a)}$$

Here $G$ and $U$ are the generation and recombination rates, respectively, $\mu_n(\mu_p)$ the electrons (holes) mobilities, $D_n$ ($D_p$) the electrons (holes) diffusion coefficients and $\xi$ is the electric field. The latter parameter could be neglected under certain circumstances, but in general it should be calculated after the Poisson's equation

$$\varepsilon\varepsilon_0 \nabla\xi = -\nabla^2\varphi = q[p - n + N_{ion}] \qquad \text{(S9.b)}$$

where $\varepsilon$ is the dielectric constant, $\varepsilon_0$ is the vacuum permittivity, $\varphi$ the electrostatic potential and $N_{ion}$ relates the total amount of ionic charge, including fixed ionized doping levels of donors $N_D$, acceptors $N_A$ and/or other like fixed ionized deep trap levels or mobile ions. Knowing the $n$, $p$, and $\xi$ allows to calculate the electron and holes current densities

$$J_n = q \, [\mu_n \, n \, \xi + D_n \, \nabla n]$$
$$J_p = q \, [\mu_p \, p \, \xi - D_p \, \nabla p] \qquad \text{(S9.c)}$$

which together compose the total current $J = J_n + J_p$.

These equations should fulfill the constancy of total current $J$ and the continuity of the electric displacement $\mathfrak{D} = \varepsilon\varepsilon_0\xi$ within the electrodes. At the electrodes, the potential and the current should be defined. The external voltage and the Galvani potential difference may be related as

$$\varphi(0) = 0,\ \ \varphi(L) = -\int_0^L \xi\, dx = V - V_{bi} \tag{S10.a}$$

where one electrode is set at position $x = 0$ and the other at $x = L$ and $V_{bi}$ is the built-in voltage. The current at the electrodes depend on the contact type: Schottky or Ohmic.[4] In the latter case the surface recombination effects are the typical approach, which results

$$J_n = qS_{rn}\,[n - n_0]$$
$$J_p = qS_{rp}[n - n_0] \tag{S10.b}$$

Here $S_{rn}$ and $S_{rp}$ are the surface recombination velocities for electrons and holes, respectively. In (S10.b) $S_{rn}$ and $S_{rp}$ are not necessary the same among them and at each electrode. For instance, in the perfect selectivity approximation and taking 0 and $L$ at the interfaces with the electrons and holes transport layers, ETL and HTL respectively, we obtain $S_{rn}(L) = S_{rp}(0) = 0$.

### S1.3. Boundary conditions

The current boundary condition is set at the electrodes as in (S10.b), stating negligible drift current so the diffusion current at such interfaces is proportional to the surface recombination as

$$D\,\frac{d\overline{n}}{dx}\Big|_{-L,L} \approx S_r\overline{n} \tag{S11.a}$$

$$D\,\frac{d\tilde{n}}{dx}\Big|_{-L,L} \approx S_r\tilde{n} \tag{S11.b}$$

where $D$ is the diffusion coefficient and $S_r$ is the effective surface recombination velocity. Note that when referring to minority holes the sign of the corresponding $D$ should be negative. Also note that in practice there is an individual $S_r$ for each carrier type at each interface, only that here for simplicity we take the average or the relevant minority carrier one.

For the potential boundary condition, note that the connection between the distribution of charge across the junction and the electrostatic potential -mirrored by the intrinsic energy level $E_i$- is usefully expressed in terms of the quasi-fermi levels of electrons $E_{Fn}$ and holes $E_{Fp}$. In our deductions we take $n$ as an average minority carrier density when the space gradients can be neglected or one of the minority carriers at quasi-neutral region of the selective contacts in the case of one-sided abrupt p-n junction. For instance, in the p-type quasi-neutral region the minority electrons are modulated as $E_{Fn}$

$$n = n_i \exp\left[\frac{E_{Fn} - E_i}{k_B T}\right] \tag{S12.a}$$

While the concentration of majority carriers is nearly constant below high injection

$$p = n_i \exp\left[\frac{E_i - E_{Fp}}{k_B T}\right] \cong N_A \tag{S12.b}$$

here $n_i$ is the intrinsic carrier density and at the border of the depletion region towards the quasi-neutral zone

$$qV = E_{Fn} - E_{Fp} \tag{S12.c}$$

From (S12) it results that Equation (S2) can be rewritten at a boundary of the space charge region as

$$n(w) \approx n_0 \left( \exp\left[\frac{qV}{k_B T}\right] - 1 \right) \tag{S13}$$

Equation (S13) basically illustrates the way the minority charge carriers increase with bias above the equilibrium level at the border of the space charge region, as schemed in **Figure S1**. In that figure, it is highlighted how in this approximation the short circuit regime may complement the forward bias that match the corresponding open circuit, only with the space shift due to the shrinking of the depletion zone, which follows the applied voltage as

$$w \cong \sqrt{\frac{2\varepsilon_0 \varepsilon}{q\, N_{D,A}} (V_{bi} - V)} \tag{S14}$$

Where $\varepsilon_0$ is the vacuum permittivity, $\varepsilon$ the dielectric constant, $V_{bi}$ the built-in voltage and $N_{D,A}$ is the lowest of the donor/acceptor doping densities in the one-side abrupt p-n junction approximation, or the result of the inverse adding.

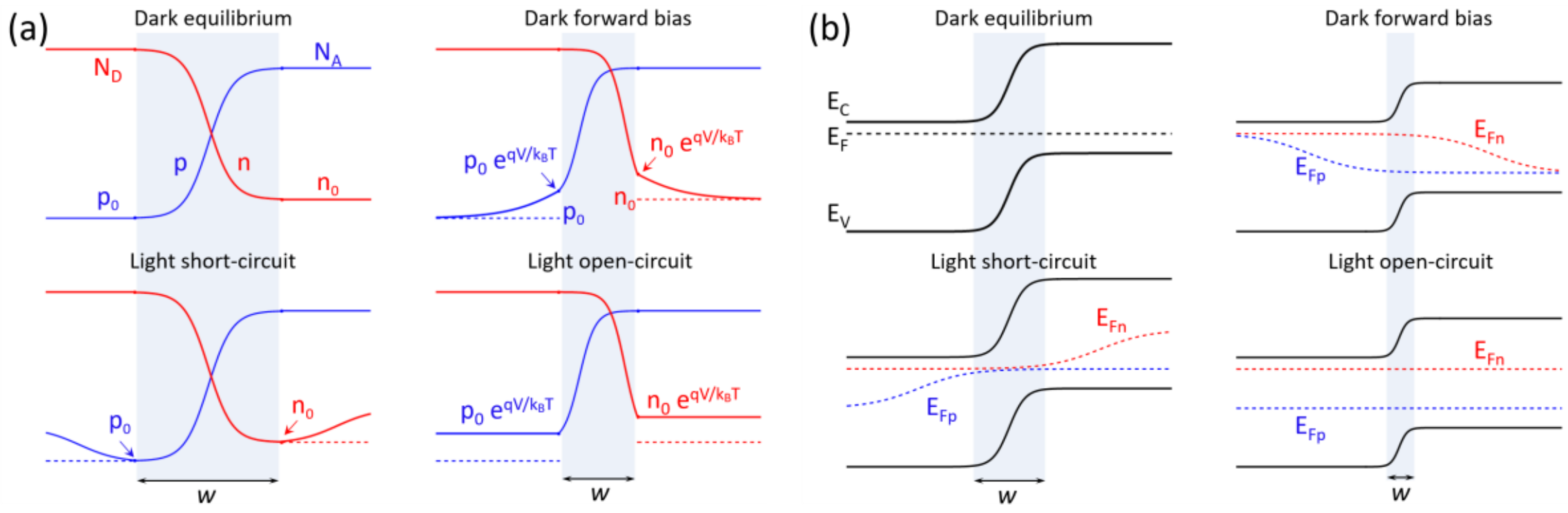

**Figure S1.** Typical steady-state p-n junction carrier concentrations (a) and corresponding energy diagrams (b) for different situations, as indicated.

### S1.4. Open circuit *dc* problem

The focus of our analysis is the OC regime, which makes a good situation for using the abode introduced average $n$. By doing so we should consider that the validity of our results depends on the accuracy of the low injection regime, where the field dependent drift contributions can be neglected. Said this, the continuity equation under illumination can be approached for our average carrier density constant charge density to

$$\frac{dn}{dt} = G - \frac{n}{\tau} - \beta n^2 + D\frac{d^2 n}{dx^2} \tag{S15}$$

Where $G$ is the generation rate, $\tau$ is the surface/SRH non-radiative recombination lifetime, $\beta$ is the radiative recombination coefficient and $D$ is the diffusion coefficient. Note that here we consider the main "ingredients" for the simulation of a rectifying semiconductor device.

By substituting Equation (S2) into (S15) and considering $\tilde{n} \ll \overline{n}_0$ we can approach to

$$i\omega\, \tilde{n} \exp[i\omega t] = G - \frac{\overline{n}_0 + \tilde{n}\exp[i\omega t]}{\tau} - \beta \overline{n}_0{}^2 \left(1 + 2\frac{\tilde{n}}{\overline{n}_0} \exp[i\omega t]\right) + D\frac{d^2(\overline{n}_0 + \tilde{n}\exp[i\omega t])}{dx^2} \tag{S16}$$

Thus, the problem here is divided in two. First finding the steady-state solution ($\omega = 0$) at OC, when $\bar{J} = 0$ and the average charge distribution should be nearly space constant ($\nabla n = 0$), meaning that the diffusion term should be neglected, which lead us to

$$\beta \overline{n}_0{}^2 + \frac{\overline{n}_0}{\tau} - \bar{G} = 0 \tag{S17}$$

where $\bar{G}$ is the *dc* part of the generation rate.

Second, the modulated solution may be obtained from

$$D\frac{d^2\tilde{n}}{dx^2} - \tilde{n}\left(i\omega + \frac{1}{\tau} + 2\,\beta\,\bar{n}_0\right) + \tilde{G} = 0 \quad \text{(S18)}$$

where $\tilde{G}$ is the *ac* part of the generation rate when it is present. Here note that (S18) is a second order non-linear differential equation without solution in terms of elementary functions. Accordingly, the appropriate approximations will be made.

The steady-state problem in (S17) consist on a quadratic equation with exact solution and binomial series approximation as

$$\bar{n}_0 = \frac{1}{2\beta\tau}\left(\sqrt{1 + 4\bar{G}\beta\tau^2} - 1\right) = \frac{1}{2\beta}\left(\omega_0 - \frac{1}{\tau}\right) \approx \tau\bar{G} \quad \text{(S19.a)}$$

Where we use the characteristic generation-recombination frequency and McLaurin series approximation as

$$\omega_0 = \frac{\sqrt{1 + 4\bar{G}\beta\tau^2}}{\tau} \approx \frac{1 + 2\bar{G}\beta\tau^2}{\tau} \quad \text{(S19.b)}$$

Here note that by discarding the corresponding term directly in (S17), for predominant non-radiative recombination ($\beta \to 0$) we obtain the same approximation. However, for predominant radiative recombination $\tau \to \infty$, we define $\omega_\beta \approx 2\sqrt{\bar{G}\beta}$ , and thus

$$\bar{n}_0 \approx \sqrt{\frac{\bar{G}}{\beta}} = \frac{\omega_\beta}{2\beta} \quad \text{(S19.c)}$$

More importantly, in all the cases there is an evident proportionality with the generation rate.

### S1.5. IS: *ac* current problem

The problem of the modulated carrier density for potentiostatic IS requires $\tilde{G} = 0$ in (S18), and by considering $2\bar{n}_0 \gg \tilde{n}$ it can be re-written as

$$\frac{d^2\tilde{n}}{dx^2} - \frac{\omega_0\left(1 + i\frac{\omega}{\omega_0}\right)}{D}\tilde{n} = 0 \qquad \text{(S20)}$$

or in the case of predominant radiative recombination using (S19.c) it will result the same but changing $\omega_0$ by $\omega_\beta$ in (S20). The general solution of (S20) is

$$\tilde{n} = \tilde{n}_1 \exp\left[-\frac{x}{\tilde{L}_d}\right] + \tilde{n}_2 \exp\left[\frac{x}{\tilde{L}_d}\right] \qquad \text{(S21.a)}$$

with the complex diffusion length as

$$\tilde{L}_d = \sqrt{\frac{D}{\omega_0}\frac{1}{\left(1+i\frac{\omega}{\omega_0}\right)}} \qquad \text{(S21.b)}$$

and the space constants $\tilde{n}_1$ and $\tilde{n}_2$ defined by the boundary conditions.

Applying our first constriction (S13) implies modulation of the depletion layer width and minority carrier concentrations at its border. By substituting (S3) in (S14) and taking the binomial series approximation for $\tilde{V} \ll (V_{bi} - \bar{V})$ the *dc* and *ac* components of the depletion layer can be approached as

$$\bar{w} \approx \sqrt{\frac{2\varepsilon_0\varepsilon}{q\,N_{D,A}}(V_{bi} - \bar{V})} \qquad \text{(S22.a)}$$

$$\tilde{w} \approx -\sqrt{\frac{\varepsilon_0\varepsilon}{q\,N_{D,A}2(V_{bi} - \bar{V})}}\tilde{V} \qquad \text{(S22.b)}$$

Note that (S22.b) loses validity as $\bar{V} \to V_{bi}$ and the minus sign expresses that $\tilde{w}$ shrinks when $\tilde{V}$ increases. Accordingly, in the one side abrupt junction case the condition (S13) may be evaluated for $x = \overline{w}$. Hence, by substituting (S3) in (S13), considering the forward bias low injection range with $\overline{V} > 4k_BT/q$, and using the McLaurin series approximation, we can get an expression for the average charge carrier concentration with the same structure of (S2) where

$$\overline{n}_0(\overline{w}) \cong n_0 \exp\left[\frac{q\overline{V}}{k_BT}\right] \tag{S23.a}$$

$$\tilde{n}(\overline{w}) \approx \overline{n}_0 \frac{q\tilde{V}}{k_BT} \tag{S23.b}$$

Here, first note that relation (S23.a) approaches (S19) when $\overline{V}$ is that of the OC regime.

For our second boundary condition, by substituting (S21.a) in (S11.b) it results

$$\frac{D}{\tilde{L}_d}\left(-\tilde{n}_1 \exp\left[-\frac{L}{\tilde{L}_d}\right] + \tilde{n}_2 \exp\left[\frac{L}{\tilde{L}_d}\right]\right) = S_r\left(\tilde{n}_1 \exp\left[-\frac{L}{\tilde{L}_d}\right] + \tilde{n}_2 \exp\left[\frac{L}{\tilde{L}_d}\right]\right) \tag{S23.c}$$

The solution of the system gives

$$\tilde{n} \cong \overline{n}_0 \frac{q\tilde{V}}{k_BT}\left(\frac{\left(1+\frac{\tilde{L}_dS_r}{D}\right)e^{\frac{x}{\tilde{L}_d}} + \left(1-\frac{\tilde{L}_dS_r}{D}\right)e^{\frac{2L-x}{\tilde{L}_d}}}{e^{\frac{2L-\overline{w}}{\tilde{L}_d}} + e^{\frac{\overline{w}}{\tilde{L}_d}} + \frac{\tilde{L}_dS_r}{D}\left(e^{\frac{\overline{w}}{\tilde{L}_d}} - e^{\frac{2L-\overline{w}}{\tilde{L}_d}}\right)}\right) \tag{S24.a}$$

Particularly, for near flat-band ($\overline{w} \ll \overline{L}_d$) and high mobility thin film devices ($2L \ll \overline{L}_d$) the parentheses in (S24.a) approaches $\cosh[x/\tilde{L}_d] + \sinh[x/\tilde{L}_d]\tilde{L}_dS_r/D$. In the assumption of the one side abrupt n$^{++}$-p junction with negligible diffusion length in the n-side, the integral of (S24.a) between 0 and $L$ gives

$$\int \tilde{n}\,dx = \overline{n}_0 \frac{q\tilde{V}}{k_BT}\tilde{L}_d\left(\frac{\left(e^{\frac{L}{\tilde{L}_d}}-1\right)\left(1+\frac{\tilde{L}_dS_r}{D}+\left(1-\frac{\tilde{L}_dS_r}{D}\right)e^{\frac{L}{\tilde{L}_d}}\right)}{e^{\frac{2L-\overline{w}}{\tilde{L}_d}} + e^{\frac{\overline{w}}{\tilde{L}_d}} + \frac{\tilde{L}_dS_r}{D}\left(e^{\frac{\overline{w}}{\tilde{L}_d}} - e^{\frac{2L-\overline{w}}{\tilde{L}_d}}\right)}\right) \tag{S24.b}$$

Thus we can take the parentheses of (S24.b) as

$$\tilde{\gamma} = \frac{\left(e^{\frac{L}{\tilde{L}_d}} - 1\right)\left(1 + \frac{\tilde{L}_d S_r}{D} + \left(1 - \frac{\tilde{L}_d S_r}{D}\right) e^{\frac{L}{\tilde{L}_d}}\right)}{e^{\frac{2L-\overline{w}}{\tilde{L}_d}} + e^{\frac{\overline{w}}{\tilde{L}_d}} + \frac{\tilde{L}_d S_r}{D}\left(e^{\frac{\overline{w}}{\tilde{L}_d}} - e^{\frac{2L-\overline{w}}{\tilde{L}_d}}\right)} \tag{S24.c}$$

Subsequently, we can evaluate $\tilde{n}$ in (S5.b) and then substituting in (S1) it results

$$Z \cong \frac{\tilde{V}}{q\tilde{L}_d \frac{q\tilde{V}\tilde{\gamma}}{k_B T} \frac{\left(\omega_0 - \frac{1}{\tau}\right)}{2\beta} (\omega_0 + i\omega)} \cong \frac{2k_B T\, \beta}{q^2 \tilde{L}_d \left(\omega_0{}^2 - \frac{\omega_0}{\tau}\right)} \frac{1}{\tilde{\gamma}\left(1 + i\frac{\omega}{\omega_0}\right)}$$

$$Z \approx \frac{2k_B T}{q^2 \overline{L}_d \overline{G}} \frac{1}{\tilde{\gamma}\sqrt{1 + i\frac{\omega}{\omega_0}}} \tag{S25}$$

### S1.6. IMVS: *ac* voltage problem

In the case of light modulation at OC, we take the steady-state *dc* solution of the average charge density $\overline{n}_0$ as Equation (S19), which is the best approximation for the IMVS. On the other hand, for the *ac* solution the problem should be split again, in solving first the transfer function for the IMVS and later with the IMPS.

In the case of finding the photo-voltage solution, the OC condition makes plausible to neglect the diffusion term in Equation (S18). Hence, we can discard the quadratic term ($2\overline{n}_0 \gg \tilde{n}$) and the solution and binomial series approximation are

$$\tilde{n} \cong \frac{\tilde{G}}{\omega_0} \frac{1}{\left(1 + i\frac{\omega}{\omega_0}\right)} \approx \frac{\tilde{G}\tau}{(1 + 2\overline{G}\beta\tau^2)} \frac{1}{\left(1 + i\frac{\omega}{\omega_0}\right)} \tag{S26.a}$$

with $\omega_0 = \sqrt{1 + 4\overline{G}\beta\tau^2}/\tau$, or in the case of predominant radiative recombination

$$\tilde{n} \approx \frac{\tilde{G}}{\omega_\beta} \frac{1}{\left(1 + i\frac{\omega}{\omega_\beta}\right)} = \frac{\tilde{G}}{2\sqrt{\overline{G}\beta}} \frac{1}{\left(1 + i\frac{\omega}{\omega_\beta}\right)} \tag{S26.b}$$

with $\omega_\beta = 2\sqrt{\bar{G}\beta}$. Subsequently, these solutions can be substituted in Equation (S8.b) so the *ac* photo-voltage can be found and approximated as

$$\tilde{V}_{oc} \cong 2\frac{k_B T}{q}\frac{2\beta\tau}{\omega_0 - \frac{1}{\tau}}\frac{\tilde{G}}{\sqrt{1+4\bar{G}\beta\tau^2}}\frac{1}{\left(1+i\frac{\omega}{\omega_0}\right)}$$

$$\tilde{V}_{oc} \cong 4\frac{k_B T}{q}\frac{\tilde{G}\beta}{({\omega_0}^2 - \omega_0/\tau)}\frac{1}{\left(1+i\frac{\omega}{\omega_0}\right)} \approx 2\frac{k_B T}{q}\frac{\tilde{G}}{\bar{G}}\frac{1}{\left(1+i\frac{\omega}{\omega_0}\right)} \tag{S27.a}$$

or in the case of predominant radiative recombination

$$\tilde{V}_{oc} \cong 2\frac{k_B T}{q}\frac{1}{\sqrt{\bar{G}/\beta}}\frac{\tilde{G}}{2\sqrt{\bar{G}\beta}}\frac{1}{\left(1+i\frac{\omega}{\omega_\beta}\right)}$$

$$\tilde{V}_{oc} \approx \frac{k_B T}{q}\frac{\tilde{G}}{\bar{G}}\frac{1}{\left(1+i\frac{\omega}{\omega_\beta}\right)} \tag{S27.b}$$

Here we see that the coefficient of the ac signal approaches 2 or 1 when significant or negligible SRH recombination, respectively, which allow us to justify the low frequency limit

$$\tilde{V}_{oc} = m\frac{k_B T}{q}\frac{\tilde{G}}{\bar{G}} \tag{S27.c}$$

being $m$ the typical ideality factor often taken in the empirical approximation to the Shockley equation. With the information of the *ac* photovoltage, then we can substitute (S27) in the definition of voltage responsivity $\Psi_V = \tilde{V}_{oc}/\tilde{P}_{in}$ resulting

$$\Psi_V(\omega) \cong 2\frac{k_B T\,\Psi_{sc}}{q^2 L\bar{G}}\frac{1}{\left(1+i\frac{\omega}{\omega_0}\right)} = 2\frac{k_B T}{q\,\bar{P}_{in}}\frac{1}{\left(1+i\frac{\omega}{\omega_0}\right)} \tag{S28.a}$$

for significant SRH recombination. Here $\Psi_{sc}$ is the photo-current responsivity at short-circuit that depends on the incident light spectrum, the absorption coefficient and the geometry of the absorbing materials. Moreover, for predominant radiative recombination

$$\Psi_V(\omega) \cong \frac{k_B T\, \Psi_{sc}}{q^2 L \bar{G}} \frac{1}{\left(1 + i\frac{\omega}{\omega_0}\right)} = \frac{k_B T}{q\, \bar{P}_{in}} \frac{1}{\left(1 + i\frac{\omega}{\omega_0}\right)} \tag{S28.b}$$

and in the low frequency limit

$$\Psi_V = m\frac{k_B T\, \Psi_{sc}}{q^2 L \bar{G}} = m\frac{k_B T}{q\, \bar{P}_{in}} \tag{S28.c}$$

Importantly, note that in the low frequency limit trend $\Psi_V \propto \bar{G}^{-1}$ is a very useful result in order to validate our calculus and, in addition, serves as an evaluation technique for the ideality factor. Towards higher frequencies $\left(1 + i\frac{\omega}{\omega_0}\right)^{-1}$ delivers a typical arc-shape-type spectrum.

### S1.7. IMPS: *ac* current problem

Moreover, regarding the IMPS the *ac* solution comes by setting Equation (S20) equal to $\tilde{G}$, whose general solution is that of (S21) plus the term $\tilde{G}/\omega_0(1 + i\omega/\omega_0)$. For the determination of the constants the first boundary condition fixes the carrier density at the border of the depletion region to the *dc* value:

$$\tilde{n}(\bar{w}) \approx 0 \tag{S29.a}$$

This approximation illustrates the limit difference between IS and LIMIS. Later, the surface recombination constriction at the electrodes is written here as

$$\frac{D}{L_D}\left(-\tilde{n}_1\exp\left[-\frac{L}{\tilde{L}_d}\right]+\tilde{n}_2\exp\left[\frac{L}{\tilde{L}_d}\right]\right)$$

$$=S_r\left(\tilde{n}_1\exp\left[-\frac{L}{\tilde{L}_d}\right]+\tilde{n}_2\exp\left[\frac{L}{\tilde{L}_d}\right]+\frac{\tilde{G}}{\omega_0(1+i\omega/\omega_0)}\right) \tag{S29.b}$$

After substitution of the solved constants on the general solution, the result is

$$\tilde{n}\cong\frac{\tilde{G}}{\omega_0(1+i\omega/\omega_0)}\left(1-\frac{\left(1+\frac{\tilde{L}_dS_r}{D}\right)e^{\frac{x}{\tilde{L}_d}}+\left(1-\frac{\tilde{L}_dS_r}{D}\right)e^{\frac{2L-x}{\tilde{L}_d}}}{e^{\frac{2L-\bar{w}}{\tilde{L}_d}}+e^{\frac{\bar{w}}{\tilde{L}_d}}+\frac{\tilde{L}_dS_r}{D}\left(e^{\frac{\bar{w}}{\tilde{L}_d}}-e^{\frac{2L-\bar{w}}{\tilde{L}_d}}\right)}\right.$$
$$\left.+\frac{\frac{\tilde{L}_dS_r}{D}\left(e^{\frac{L+\bar{w}-x}{\tilde{L}_d}}-e^{\frac{L-\bar{w}+x}{\tilde{L}_d}}\right)}{e^{\frac{2L-\bar{w}}{\tilde{L}_d}}+e^{\frac{\bar{w}}{\tilde{L}_d}}+\frac{\tilde{L}_dS_r}{D}\left(e^{\frac{\bar{w}}{\tilde{L}_d}}-e^{\frac{2L-\bar{w}}{\tilde{L}_d}}\right)}\right) \tag{S30.a}$$

Similarly to (S24.b) the integral of (S30.a) between 0 and $L$ gives

$$\int\tilde{n}\,dx=\frac{\tilde{G}}{\omega_0(1+i\omega/\omega_0)}\tilde{L}_d\left(\frac{L}{\tilde{L}_d}-\tilde{\gamma}\right.$$
$$\left.+\frac{\frac{\tilde{L}_dS_r}{D}e^{-\frac{\bar{w}}{\tilde{L}_d}}\left(e^{\frac{L}{\tilde{L}_d}}-1\right)\left(e^{\frac{2\bar{w}}{\tilde{L}_d}}-e^{\frac{L}{\tilde{L}_d}}\right)}{e^{\frac{2L-\bar{w}}{\tilde{L}_d}}+e^{\frac{\bar{w}}{\tilde{L}_d}}+\frac{\tilde{L}_dS_r}{D}\left(e^{\frac{\bar{w}}{\tilde{L}_d}}-e^{\frac{2L-\bar{w}}{\tilde{L}_d}}\right)}\right) \tag{S30.b}$$

Thus we can take the last term in the parentheses of (S30.b) as

$$\tilde{\gamma}_1=\frac{\frac{\tilde{L}_dS_r}{D}e^{-\frac{\bar{w}}{\tilde{L}_d}}\left(e^{\frac{L}{\tilde{L}_d}}-1\right)\left(e^{\frac{2\bar{w}}{\tilde{L}_d}}-e^{\frac{L}{\tilde{L}_d}}\right)}{e^{\frac{2L-\bar{w}}{\tilde{L}_d}}+e^{\frac{\bar{w}}{\tilde{L}_d}}+\frac{\tilde{L}_dS_r}{D}\left(e^{\frac{\bar{w}}{\tilde{L}_d}}-e^{\frac{2L-\bar{w}}{\tilde{L}_d}}\right)} \tag{S30.c}$$

Subsequently, we can evaluate $\tilde{n}$ from (S30) in (S5.b) considering the fact that the *ac* current will always be negative in our representation, then

$$\tilde{J} \approx -q\tilde{L}_d\tilde{G}\left(\frac{\left(\frac{L}{\tilde{L}_d}-\tilde{\gamma}+\tilde{\gamma}_1\right)}{\omega_0\left(1+i\frac{\omega}{\omega_0}\right)}\left(\frac{1}{\tau}+\frac{2\beta}{2\beta\tau}\left(\sqrt{1+4\bar{G}\beta\tau^2}-1\right)+i\omega\right)-\frac{L}{\tilde{L}_d}\right)$$

$$\tilde{J} \approx \frac{q\bar{L}_d\tilde{G}}{\sqrt{1+i\frac{\omega}{\omega_0}}}\left(\frac{\left(\tilde{\gamma}-\tilde{\gamma}_1-\frac{L}{\tilde{L}_d}\right)}{\omega_0\left(1+i\frac{\omega}{\omega_0}\right)}\omega_0\left(1+i\frac{\omega}{\omega_0}\right)+\frac{L}{\tilde{L}_d}\right) \quad \text{(S31)}$$

$$\tilde{J} \approx \frac{q\bar{L}_d\tilde{G}}{\sqrt{1+i\frac{\omega}{\omega_0}}}\tilde{\gamma}\left(1-\tilde{\delta}\right)$$

Where $\tilde{\delta} = \tilde{\gamma}_1/\tilde{\gamma}$ is

$$\tilde{\delta} \cong \frac{\frac{\tilde{L}_d S_r}{D}\left(e^{\frac{\bar{w}}{\tilde{L}_d}}-e^{\frac{L-\bar{w}}{\tilde{L}_d}}\right)}{\left(1+\frac{\tilde{L}_d S_r}{D}+\left(1-\frac{\tilde{L}_d S_r}{D}\right)e^{\frac{L}{\tilde{L}_d}}\right)} \quad \text{(S32)}$$

Here, given the information about the photocurrent, the current responsivity can be obtained by substituting (S31) in the definition (6) from the main manuscript, resulting

$$\Psi_J \approx \frac{q\bar{L}_d}{\sqrt{1+i\frac{\omega}{\omega_0}}}\tilde{\gamma}\left(1-\tilde{\delta}\right) \quad \text{(S33)}$$

Importantly, note that in the low frequency limit $\Psi_J$ is nearly independent on the dc illumination intensity. This is a clear difference with respect to the photovoltage responsivity and a useful argument to validate our theoretical approximations. Towards higher frequencies $\left(1+i\frac{\omega}{\omega_0}\right)^{-1/2}$ delivers a typical arc-shape-type spectrum.

### S1.8. LIMIS

Finally, the impedance definition (S1) can be rewritten as (9) in the main manuscript, by substituting the *ac* photovoltage from (S27) and the *ac* photocurrent from (S31) which results

$$Z_\Psi \approx \frac{2k_B T\tilde{G}}{q\bar{G}\left(1+i\frac{\omega}{\omega_0}\right)} \frac{1}{\frac{q\bar{L}_d\tilde{G}}{\sqrt{1+i\frac{\omega}{\omega_0}}}\tilde{\gamma}\left(1-\tilde{\delta}\right)}$$

$$Z_\Psi \approx \frac{2k_B T}{q^2\bar{L}_d\bar{G}} \frac{1}{\tilde{\gamma}\sqrt{1+i\frac{\omega}{\omega_0}}\left(1-\tilde{\delta}\right)} \tag{S34}$$

# S2. IS-LIMIS difference and the simulations

$$\Delta Z_\Psi = \frac{Z_\Psi - Z}{Z} = \frac{\tilde{\delta}}{1-\tilde{\delta}}$$

$$= \frac{\frac{\tilde{L}_d S_r}{D}\left(e^{\frac{\bar{w}}{\tilde{L}_d}} - e^{\frac{L-\bar{w}}{\tilde{L}_d}}\right)}{\left(1+\frac{\tilde{L}_d S_r}{D} + \left(1-\frac{\tilde{L}_d S_r}{D}\right)e^{\frac{L}{\tilde{L}_d}}\right)} \frac{1}{1-\frac{\frac{\tilde{L}_d S_r}{D}\left(e^{\frac{\bar{w}}{\tilde{L}_d}} - e^{\frac{L-\bar{w}}{\tilde{L}_d}}\right)}{\left(1+\frac{\tilde{L}_d S_r}{D} + \left(1-\frac{\tilde{L}_d S_r}{D}\right)e^{\frac{L}{\tilde{L}_d}}\right)}} \tag{S35}$$

$$\Delta Z_\Psi = \frac{\frac{\tilde{L}_d S_r}{D}\left(e^{\frac{L}{\tilde{L}_d}} - e^{\frac{2\bar{w}}{\tilde{L}_d}}\right)}{\left(\frac{\tilde{L}_d S_r}{D}\left(e^{\frac{\bar{w}}{\tilde{L}_d}} - 1\right)\left(e^{\frac{L}{\tilde{L}_d}} + e^{\frac{\bar{w}}{\tilde{L}_d}}\right)\right) - e^{\frac{\bar{w}}{\tilde{L}_d}} - e^{\frac{L+\bar{w}}{\tilde{L}_d}}}$$

**Table S2:** Simulation parameters for Figure 2 in the main manuscript. The "*" signals the values as measured in the experiment in our simultaneous work.[5]

| Parameter | Values per graph | | |
|---|---|---|---|
| | (a) | (b) | (c) |
| T (K) | 300 | | |
| $\epsilon$ | 11.68 | | |
| $N_A$ (cm$^{-3}$)* | $1.17 \times 10^{15}$ | | |
| $V_{bi}$ (mV)* | 700 | | |
| $\tilde{V}$ (mV) | 10 | | |
| $\sqrt{N_C N_V}$ (cm$^{-3}$) | $1.27 \times 10^{19}$ | | |
| $E_g$ (eV) | 1.10 | | |
| $\mu$ (cm$^2$·V$^{-1}$·s$^{-1}$) | 1000 | | |
| $L$ (nm) | 900 | | |
| $\tau$ (µs) | 31.6 | 100 | 10 |
| $\beta$ (cm$^3$·s$^{-1}$) | $6.17 \times 10^{-9}$ | $6.17 \times 10^{-9}$ | $7.94 \times 10^{-11}$ |
| $S_r$ (cm·s$^{-1}$) | $1.02 \times 10^{5}$ | $1.17 \times 10^{5}$ | $1.00 \times 10^{5}$ |
| $\bar{G}$ (cm$^{-3}$·s$^{-1}$) | $4.36 \times 10^{19}$ | $7.94 \times 10^{18}$ | $4.68 \times 10^{20}$ |
| $R_s$ (Ω·cm$^2$) | 22 | 17 | 3.65 |
| $Z_s{}'$ (Ω·cm$^2$) | 0 | 88 | 0.8 |
| $\bar{L}_d$ (µm) | 49.9 | 76.4 | 80.6 |
| $\bar{w}$ (nm) | 523 | 568 | 308 |
| $\bar{V}_{oc}$ (mV) | 451 | 408 | 614 |
| $\bar{V}_{oc}$ (mV)* | 386 | 322 | 467 |
| $\bar{P}_{in}$ (mV)* | 20 | 5 | 200 |

# Author Information

Corresponding Author

*E-mail: osbel.almora@fau.de, almora@uji.es

# ORCID

Osbel Almora: 0000-0002-2523-0203

Daniel Miravet: 0000-0002-2908-4645